\documentclass[]{sig-alternate-05-2015}
\usepackage{lmodern}
\usepackage{amssymb,amsmath}
\usepackage{ifxetex,ifluatex}
\usepackage{fixltx2e} 
\ifnum 0\ifxetex 1\fi\ifluatex 1\fi=0 
  \usepackage[T1]{fontenc}
  \usepackage[utf8]{inputenc}
\else 
  \ifxetex
    \usepackage{mathspec}
  \else
    \usepackage{fontspec}
  \fi
  \defaultfontfeatures{Ligatures=TeX,Scale=MatchLowercase}
\fi
\IfFileExists{upquote.sty}{\usepackage{upquote}}{}
\IfFileExists{microtype.sty}{%
\usepackage{microtype}
\UseMicrotypeSet[protrusion]{basicmath} 
}{}
\usepackage{hyperref}
\hypersetup{unicode=true,
            pdftitle={Attribution of Responsibility and Blame Regarding a Man-made Disaster: \#FlintWaterCrisis},
            pdfborder={0 0 0},
            breaklinks=true}
\IfFileExists{parskip.sty}{%
\usepackage{parskip}
}{
\setlength{\parindent}{0pt}
\setlength{\parskip}{6pt plus 2pt minus 1pt}
}
\ifx\paragraph\undefined\else
\let\oldparagraph\paragraph
\renewcommand{\paragraph}[1]{\oldparagraph{#1}\mbox{}}
\fi
\ifx\subparagraph\undefined\else
\let\oldsubparagraph\subparagraph
\renewcommand{\subparagraph}[1]{\oldsubparagraph{#1}\mbox{}}
\fi
\usepackage{floatrow} 
\usepackage[font=bf,skip=\baselineskip,figurename={Figure},tablename={Table},listfigurename={listfigure},listtablename={listtable}]{caption}
\usepackage{booktabs}

\setcopyright{acmcopyright} \doi{00.000/000_0}
\isbn{000-0000-00-000/00/00}
\conferenceinfo{CIKM-SWDM '16}{October 24--28, 2016, Indianapolis, IN, USA}
\acmPrice{\$00.00} \numberofauthors{2} \author{
\alignauthor
Talha Oz\titlenote{Research by the author is supported by the George Mason University Center for Social Complexity and the Defense Threat Reduction Agency (DTRA) grant HDTRA1-16-0043.}\\
       \affaddr{Center for Social Complexity}\\
       \affaddr{George Mason University}\\
       \affaddr{Fairfax, VA 22030 USA}\\
       \email{toz@gmu.edu}
\alignauthor
Halil Bisgin\titlenote{Corresponding author.}\\
\affaddr{Computer Science, Engineering, and Physics}\\
\affaddr{University of Michigan-Flint}\\
\affaddr{Flint, MI 48502 USA}\\
\email{bisgin@umich.edu}
}
\usepackage{subfig}
\AtBeginDocument{%

}
\AtBeginDocument{%

}
\usepackage{float}
\floatstyle{ruled}
\makeatletter
\@ifundefined{c@chapter}{\newfloat{codelisting}{h}{lop}}{\newfloat{codelisting}{h}{lop}[chapter]}
\makeatother
\floatname{codelisting}{Listing}

\title{Attribution of Responsibility and Blame Regarding a Man-made Disaster:
\#FlintWaterCrisis}
\date{}

\begin{document}
\maketitle
\begin{abstract}
Attribution of responsibility and blame are important topics in
political science especially as individuals tend to think of political
issues in terms of questions of responsibility, and as blame carries far
more weight in voting behavior than that of credit. However,
surprisingly, there is a paucity of studies on the attribution of
responsibility and blame in the field of disaster research.

The Flint water crisis is a story of government failure at all levels.
By studying microblog posts about it, we understand how citizens assign
responsibility and blame regarding such a man-made disaster online. We
form hypotheses based on social scientific theories in disaster research
and then operationalize them on unobtrusive, observational social media
data. In particular, we investigate the following phenomena: the source
for blame; the partisan predisposition; the concerned geographies; and
the contagion of complaining.

This paper adds to the sociology of disasters research by exploiting a
new, rarely used data source (the social web), and by employing new
computational methods (such as sentiment analysis and retrospective
cohort study design) on this new form of data. In this regard, this work
should be seen as the first step toward drawing more challenging
inferences on the sociology of disasters from ``big social data''.
\end{abstract}

\section{Introduction}\label{introduction}

In the last decade of disaster research, there has been a proliferation
of studies exploiting information and communication technologies (ICT)
and computational methods for advancing emergency response. These same
means can be used to address social scientific inquiries of disaster
research. In this regard, instead of trying to solve a software
engineering or a disaster management problem, here we study the
sociology of disasters from a computational social science perspective.
By examining microblog posts on the Flint water crisis we try to
understand how citizens respond to a man-made disaster and to a
governmental failure online. In particular we are interested in
responses regarding \emph{attribution of blame and responsibility},
which usually takes place in the recovery phase of disasters. To study
it, we first construct theoretical hypotheses on top of existing social
theories, and then operationalize them on unobtrusive, observational
social media data via computational methods.

Attribution of responsibility is a key issue in political decision
making as blame carries far more weight in voting behavior than that of
credit {[}\protect\hyperlink{ref-lau_two_1985}{19}{]}. Moreover,
``individuals tend to simplify political issues by reducing them to
questions of responsibility and their issue opinions flow from their
answers to these questions''
{[}\protect\hyperlink{ref-iyengar_is_1994}{13}{]}. Besides, attributions
formed during states of national emergencies are of particular
importance, especially because these attributions become shared memories
of the entire nation and are long used as concrete examples of severity
of consequences of wrong policy decisions. Although about thirty years
ago Neal found it surprising that the process of blame was a neglected
topic in disaster research
{[}\protect\hyperlink{ref-neal_blame_1984}{22}{]}, tracing over the
citations that his paper has received to date and still not seeing any
article particularly discussing blame, made us even more worrisome. In
this paper, we contribute to this neglected field by testing theories of
attribution of blame and responsibility on the Flint water crisis using
new forms of data (the social web) and methods. In particular, we add to
the disaster research by addressing the issues of: i) sources for blame
regarding a disaster, ii) partisan predispositions in the blaming
behavior, iii) geographies that shows interest in the crisis the most,
and iv) the contagion of complaining (homophily, peer or network effect,
and selective exposure).

In the next section we first provide some background information about
the Flint water crisis. Following it, we lay out our hypotheses along
with the theories behind them. After that we describe the data and
discuss how we operationalize our hypotheses. In the fifth section, we
report and assess our results. Finally, we conclude our paper.

\section{Background}\label{sec:bg}

On Saturday, January 16, 2016, President Obama declared a federal state
of emergency for an area in Michigan affected by contaminated water and
authorized the Department of Homeland Security, Federal Emergency
Management Agency (FEMA) to ``coordinate all disaster relief efforts''
{[}\protect\hyperlink{ref-the_white_house_president_2016}{25}{]}. When
he later visited Flint, the most adversely affected city in Genesee
County, he described the water crisis as ``a man-made disaster'' that
was ``avoidable'' and ``preventable''
{[}\protect\hyperlink{ref-shear_ive_2016}{29}{]}, while not naming who
in particular were responsible.

For decades, Flint, MI used Detroit's treated sources for tap water.
However, Detroit's double digit price increases every year has
eventually made it the most expensive option
{[}\protect\hyperlink{ref-longley_report:_2011}{26}{]}, and on March 25,
2014 Flint city council approved buying water from Karegnondi Water
Authority (KWA) when it becomes active. Upon this decision the ``water
war'' started according to Detroit Water and Sewerage Department (DWSD),
and DSDW gave a notice that it would terminate its contract with Flint
in one year {[}\protect\hyperlink{ref-fonger_detroit_2013}{5},
\protect\hyperlink{ref-wright_genesee_2013}{10}{]}. Flint had to find a
temporary primary water source until KWA becomes effective, and by late
April 2014, they decided to switch to Flint River temporarily.
Reportedly the complaints about the tap water started right after this
change
{[}\protect\hyperlink{ref-flintwater_advisory_task_force_flint_2016}{9}{]}.

According to Flint Water Advisory Task Force (FWATF)
{[}\protect\hyperlink{ref-flintwater_advisory_task_force_flint_2016}{9}{]},
the following seven entities are responsible for the Flint water crisis
at various levels: Michigan Department of Environmental Quality (MDEQ),
Michigan Department of Health and Human Services (MDHHS), Michigan
Governor's Office, State-appointed emergency managers (EMs), Genesee
County Health Department's (GCHD), and U.S. Environmental Protection
Agency (EPA).

\section{Hypotheses}\label{sec:hypotheses}

In ``An Inventory of Sociological Findings'', Drabek
{[}\protect\hyperlink{ref-drabek_human_1986}{6}{]} discusses ``blame
assignation processes'' at the community-level in the disaster
reconstruction phase of his typology, in which he lists hypotheses on
three topics: (i) when blame occurs, (ii) purposes of blaming and how
they work out, and (iii) who those blamers are. Here, we build our
hypotheses on top of this existing sociology of disasters research.
Drabek also notes the scarcity of studies on blame assignment behavior
in disaster research, by forming and testing hypotheses we hope our
research helps reduce this gap of knowledge in the field.

\textbf{Source for Blame.} ``Animated by a desire for prevention of
future occurrences'', blame occurs especially when (i) conventional
explanations failed, (ii) when the responsible agents are perceived to
be unwilling to take action to remedy the situation, and (iii) when they
violate moral standards
{[}\protect\hyperlink{ref-bucher_blame_1957}{4}{]}. All of the
conditions are present in the case of Flint water crisis; (i) there is
no conventional explanation for this man-made disaster, (ii) almost all
of the agents of responsibility were reluctant to respond in time, and
(iii) the public was deprived of a basic human right, the right to safe
water. Yet, per condition (ii), every primarily responsible officer in
the state ``somehow payed the price'' by leaving their posts, but
Governor Snyder\footnote{They either resigned (e.g.~EPA officials and
  emergency managers), were fired (e.g.~the head of MDEQ's drinking
  water unit), or their effective terms ended (e.g.~the mayor).}
{[}\protect\hyperlink{ref-flintwater_advisory_task_force_flint_2016}{9}{]}.
Moreover, both Democratic presidential candidates demanded the governor
to resign. Therefore, our first hypothesis goes:

\emph{\textbf{H1.} The amount of blame directed toward Governor Snyder
exceeds any other agent.}

\textbf{Partisan Predisposition.} Blaming an entire party or an ideology
upon a particular crisis predisposes him against that party. In
disasters, sometimes blame is not seen as ``a function of the immediate
crisis, but that reflect pre-existing conflicts and hostilities'', and
when biased or irrational factors play a role in the process of blaming,
it is called ``scapegoating''
{[}\protect\hyperlink{ref-singer_introduction_1982}{31}{]} (cited in
{[}\protect\hyperlink{ref-drabek_human_1986}{6}{]}). One can relate this
to the social identity theory, which suggests that if someone is guilty
then (s)he must be among the out-group
{[}\protect\hyperlink{ref-simon_politicized_2001}{30}{]}. Theories on
partisan bias project this socio-psychological bias onto the political
plane, suggesting that partisanship has an important influence on
attitudes toward political elements
{[}\protect\hyperlink{ref-bartels_beyond_2002}{2}{]}. So we expect
people blaming a particular party or ideology to express more negative
sentiments toward representatives of that party. In our case, some of
these representatives are Democrat while others are Republican\footnote{The
  city council is made up of Democrats, the state of Michigan is ruled
  by a Republican governor, the Congress is controlled by Republicans,
  and the President is a Democrat.}, and for some, Flint poisoning is
primarily a partisan issue (e.g.
{[}\protect\hyperlink{ref-krugman_michigans_2016}{15}{]}). Hence, our
second hypothesis is:

\emph{\textbf{H2a.} Individuals who assign responsibility to the
Republican party or ideology show greater negative feeling toward the
Governor (R) than those who blame Democratic party or ideology.}

\emph{\textbf{H2b.} Individuals who assign responsibility to the
Republican party or ideology express less negative sentiment toward the
Mayor (D) than those who blame Democratic party or ideology.}

\textbf{Concerned Geographies.} Tobler's first law of geography says
``everything is related to everything else, but near things are more
related than distant things''
{[}\protect\hyperlink{ref-tobler_computer_1970}{34}{]}. In the case of
Flint water crisis, this also relates to environmental vulnerability,
suggesting that individuals who are at greater risk are more likely to
express their concerns. Flint residents are under the highest threat,
followed by the Genesee residents, followed by Michiganders. Therefore
we expect:

\emph{\textbf{H3.} Expression of concern per capita is to be the highest
for the city of Flint, followed by other cities in the Genesee county,
followed by other cities and counties in Michigan.}

\textbf{Contagion of Complaining.} Twitter is not only used as a social
network but also as a news media
{[}\protect\hyperlink{ref-kwak_what_2010}{17}{]}. In the former case,
individuals befriend with similar others (homophily) and influence each
other {[}\protect\hyperlink{ref-mcpherson_birds_2001}{21}{]}. When
Twitter serves as a news media, we expect ideological similarity between
the user and who he follows as \emph{selective exposure} suggests
{[}\protect\hyperlink{ref-sears_selective_1967}{28}{]}. Besides, one who
hears a complaint is more likely to start complaining (positive
feedback), and Kowalski offers three explanations for this in times of
disasters {[}\protect\hyperlink{ref-kowalski_complaints_1996}{14}{]}.
Accordingly:

\emph{\textbf{H4.} Individuals who express negative emotions on the
Flint water crisis have friends more negative than that of individuals
who talk more positively about the crisis.}

\begin{figure*}[htbp]
\centering
\includegraphics[width=\linewidth]{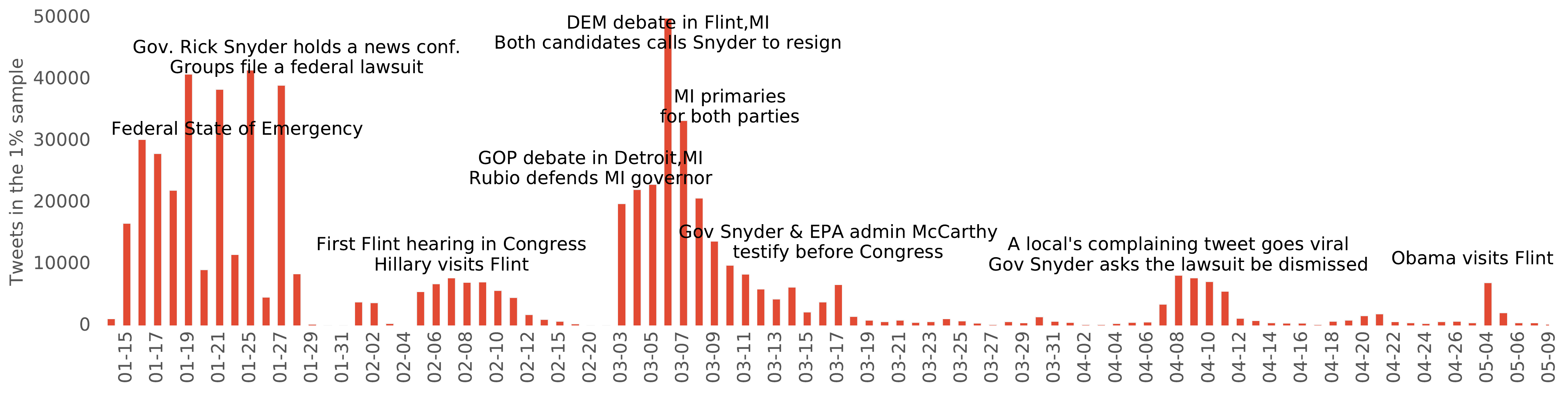}
\caption{Twitter activity on the Flint water crisis.}\label{fig:activity}
\end{figure*}

\section{Data and Methodology}\label{data-and-methodology}

One of the major advantages of social media research is that we are not
bounded with a specific space and time for data collection. This helped
us overcome a major limitation in disaster research,
\emph{unobservability}, as Wallace puts it
{[}\protect\hyperlink{ref-wallace_human_1956}{35}{]}: ``An
anthropologist can watch or participate in a religious ritual; a
sociologist can attend a union meeting; the psychiatrist can see his
patient a few hours or minutes after a family quarrel. But disasters,
generally speaking, are so unpredictable as to place and time, that it
is unlikely that any given team of trained observers will be in an
impact area, before and during an impact of the appropriate type''.
Palen et al. {[}\protect\hyperlink{ref-palen_crisis_2007}{24}{]} also
emphasize the advantages of crisis informatics in quick response
research. On the other hand, availability of big data may also obscure
the most relevant piece of information needed for an accurate conclusion
{[}\protect\hyperlink{ref-spence_social_2016}{32}{]}. To this extent, in
order to have most possible amount of data, we do not restrict ourselves
to tweets with geocoded information or that contain a particular hashtag
only. Furthermore, to not include irrelevant postings, we filter the
Twitter stream by keywords \texttt{Flint} and
\texttt{\#FlintWaterCrisis}.

We used TweetTracker
{[}\protect\hyperlink{ref-kumar_tweettracker:_2011}{16}{]} as our data
collection tool, and the data collected starts on the day before the
President declared a state of emergency for Flint. From Jan.~15 to
Jun.~29, 2016 (163 days\footnote{Data for the following days are missing
  due to collection issues: 01:23,24; 02:14,17-19; 04:28-30;
  05:1-3,7,13-25}), we obtained 664,775 tweets by 281,535 unique users.
Fig.~\ref{fig:activity} illustrates the activity on Twitter by
highlighting some of the major events that draw public's
attention\footnote{Since there is no major political event taking place
  after the President's visit on May 4th, we truncate the figure for the
  sake of better visualization}. It appears that the public interest in
the Flint water crisis has been limited, and peaked at times of major
political events. In this regard, the only day we hit the 50,000 daily
tweet collection limit of TweetTracker was the day of the Democratic
presidential debate that was held in Flint on March 6. We calculated
sentiments of the tweets in our dataset using NLTK implementation of
VADER because it is particularly designed for sentiment analysis for
social media text\footnote{VADER's sentiment lexicon includes emoticons,
  common slang words, and accounts for punctuation and capitalization.}
{[}\protect\hyperlink{ref-bird_natural_2009}{3},
\protect\hyperlink{ref-hutto_vader:_2014}{12}{]}. In the rest of this
section, we discuss how we operationalize the theoretical hypotheses put
forward in Sec.~\ref{sec:hypotheses}.

\begin{figure}[htbp]
\centering
\includegraphics[width=\columnwidth]{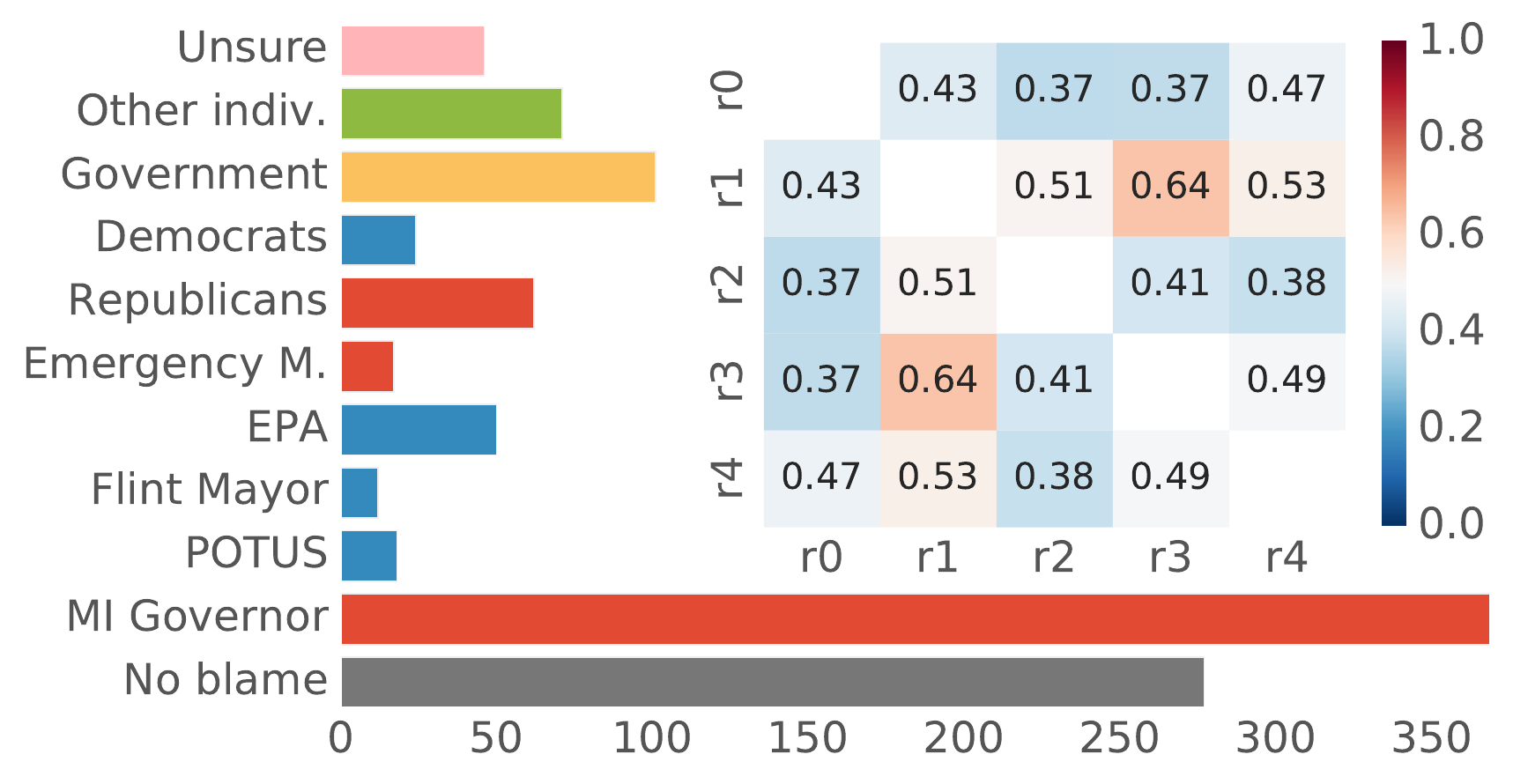}
\caption{Agents blamed and inter-rater agreement.}\label{fig:coded}
\end{figure}

\textbf{Source for Blame.} Our first hypothesis questions whether most
of the blames are directed towards Governor Snyder. To learn if a tweet,
or a phrase in a tweet, attributes blame or responsibility to any
specific person or a group, we employed manual curation. First, based on
the roles of government entities in the Flint water crisis listed in
Sec.~\ref{sec:bg} and from our preliminary observation of our dataset we
came up with eight candidates that are likely to be blamed. Then, we
randomly selected five chunks of 200 tweets from our original dataset
and asked voluntary coders\footnote{Three of the coders (Lawrence Wang,
  Varun Talwar, and Elizabeth Hu) are 2016 Summer interns in the
  Department of Computational and Data Sciences at George Mason
  University, and the other two (Feyza Galip and Kevser Polater) are
  graduate students elsewhere.} to label every tweet in a chunk with at
least one of these predefined labels (candidates). If there is no blame
attributed to any specific person or a group in a tweet, then it is
labeled \texttt{no\ blame}. If a person or group is blamed but happens
not to be in the candidates list, then those tweets are labeled as
\texttt{other}. Multiple labeling was allowed in case a tweet assigns
blame to several persons or groups. Curators were instructed not to
label a tweet if they are unsure of the person blamed, and to indicate
so. Distribution of these eleven cases is captured in
Fig.~\ref{fig:coded}. To measure inter-rater reliability, each of the
samples is created with approximately 10\% overlap with any other sample
\((\sigma =19.1, \mu =2.5)\).\footnote{Therefore, instead of 1000 tweets
  we ended up with 892 unique tweets labeled in total.} Then, to
operationalize our first hypothesis we simply evaluate the number of
tweets coded per category by the curators, for which we calculate a
Fleiss' kappa statistic for every possible coder pair. As visualized in
the heatmap in Fig.~\ref{fig:coded}, most of the rater pairs are in the
0.41--0.60 kappa range, which is interpreted as moderate
agreement\footnote{That is, \(\binom{5}{2} = 10\) pairs calculated. A
  perfect agreement would equate to a kappa of 1, and a chance agreement
  would equate to 0.}
{[}\protect\hyperlink{ref-landis_measurement_1977}{18}{]}.

\textbf{Partisan Predisposition.} Our second hypothesis is about the
relationship between explicitly blamed parties or ideologies and the
sentiments expressed toward their representatives at administrative
positions. We expect users who blame the Republican (Democratic) party
or ideology to have a more negative sentiment towards the Republican
governor (Democratic mayor) than those who blame the Democratic
(Republican) party or ideology. In our manually coded tweets sample
(Fig.~\ref{fig:coded}) two of the labels indicate tweets explicitly
blaming parties or ideologies. A total of 62 (24) of the 892 labeled
tweets found to be blaming Republicans (Democrats) for the crisis. After
identifying these tweets, we look for the individuals (Twitter accounts)
who (re)tweeted at least one of those tweets. In total, 165 such users
are identified in our main dataset, 136 of which blamed the Republicans,
and 29 blamed the Democrats.

We used keyword filtering to identify the tweets mentioning the governor
of Michigan (G), the mayor of Flint (M), and the emergency managers
(EM)\footnote{Tweets labeled for G (97577), M (11609) and E (6028) using
  keyword sets ``\texttt{governor,\ Snyder,\ onethoughnerd}'',
  ``\texttt{mayor,\ Dayne,\ Walling}'', and
  ``\texttt{mgr,\ manager,\ Kurtz,\ Earley,\ Darnell}'', respectively.},
and then selected tweets exclusively mentioning the mayor or the
governor as such: Let \(M_o:=M\setminus(G\cup EM)\) represent the set of
tweets that has only mayor-related tweets. Similarly,
\(G_o:=G\setminus(M\cup EM)\) gives the exclusively governor-related
tweets. Within each of those sets, we looked at the sentiments of
individuals blaming Republicans (R) and Democrats (D) separately. To
measure statistical difference between those who blame R and D in their
sentiments expressed toward G and M, we performed Kolmogorov-Smirnov
test, which is a non-parametric test that does not rely on any
probability distribution.

\textbf{Concerned Geographies.} Rather than working with geocoded
tweets, which are rarely available for our collection, we make use of
the \texttt{location} field in Twitter user profiles, from which we
managed to get geographic coordinates using regular expressions. Then to
measure cities' level of interest in the Flint water crisis, we
normalize total number of tweets originated at each city by its
population. For counties, we normalize total tweet counts originated
from cities in a county with the square root of sum of city populations.
We do square root transformation to account for larger standard
deviations at county level\footnote{This is due to our normalization
  factor. In normalizing Flint-related tweets per county, instead of
  using true population of counties we simply use sum of population of
  cities from which at least three tweets originated and available in
  our dataset.}.

\textbf{Contagion of Complaining.} Some Flinters have posted positive
messages about the crisis (cohort), while most others have expressed
negative sentiments (control). We expect friends of a user in any of
these two groups reflect sentiments similar to the user. To test this
hypothesis, we designed a retrospective cohort study in which we
compared the sentiments of the \texttt{friends} of the cohort group on
the Flint water crisis to that of the control group. To rule out the
geographic effect, we form both of the groups only by Flinters, the
Flinters that have at least three but no more than 20 tweets in our
dataset.\footnote{Location field in the Twitter user profile is used to
  detect the Flinters.} We found 223 such Flinters in our dataset (115
of who had a negative and 101 of who had a positive sentiment on average
on the Flint water crisis)\footnote{Following Twitter's convention, we
  use the term \texttt{friends} to refer to the users who someone
  follows. These Flinters in total follow 122,953 unique accounts, and
  8,339 of those happen to be in our dataset.}. We then performed
two-sample Kolmogorov-Smirnov statistical test that rejects the null
hypothesis if the two samples (the average sentiments of the
\texttt{friends} of each group) were drawn from the same distribution.

\section{Results}\label{results}

\textbf{Source for Blame.} As shown in Fig.~\ref{fig:coded}, the
Governor of Michigan is blamed 3.5 times more than the second most
blamed agent. Our first hypothesis expected him to be the most blamed,
and thus it is proved to be true.

\begin{figure}[htbp]
\centering
\includegraphics[width=\columnwidth]{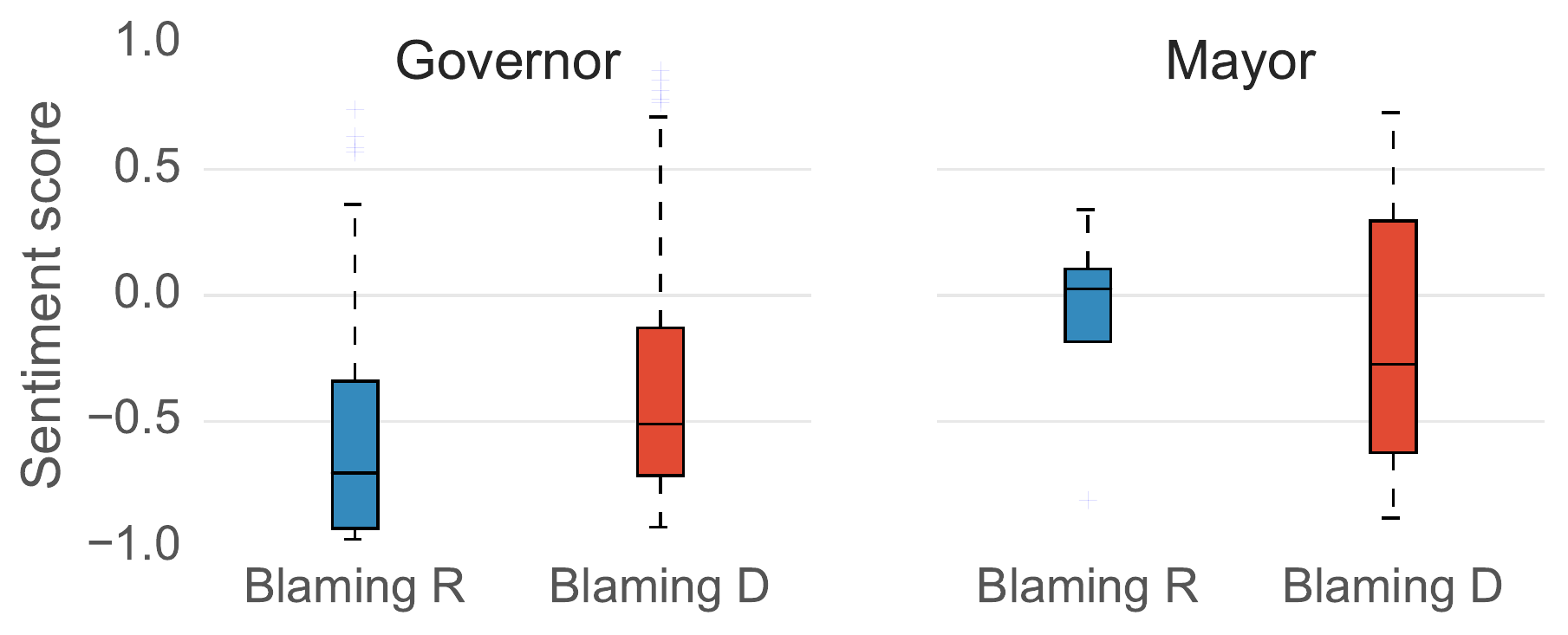}
\caption{Parties blamed and their representatives.}\label{fig:boxparty}
\end{figure}

\textbf{Partisan Predisposition.} We asked if those who blame the
Democratic party or ideology (D) is any different from those who blame
the Republican party or ideology (R) in their sentimental expressions
toward the governor, and the mayor\footnote{When we examined the
  expressions toward the mayor and the governor without separating the
  parties blamed, we found out that average sentiment scores are
  negative for both officials, though at different levels (-0.12,
  -0.31). }. Fig.~\ref{fig:boxparty} shows that individuals blaming R
have more negative sentiment toward the governor than those individuals
who blame D. The null hypothesis that the two samples (blaming R and
blaming D) are from the same distribution is rejected for the governor
by the two-sample Kolmogorov-Smirnov test (for \(\alpha=0.001\),
\(D=0.36\), p-\(value\) = \(3.5\mathrm{e}{-6}\)). Similarly,
Fig.~\ref{fig:boxparty} shows that individuals blaming R have less
negative sentiment toward the mayor than those individuals who blame D.
However, due to small sample size, we cannot statistically claim any
effect of partisan predisposition on the sentiments expressed about the
mayor. Thus, our statistical tests support \emph{H2a} but not
\emph{H2b}.

\textbf{Concerned Geographies.} We expected that the cities expressed
interest in the Flint water crisis the most to be from the county of
Genesee and from the state of Michigan. Four of the ten most concerned
cities happen to be from the county of Genesee, and six of the ten
counties are from Michigan (Tbl.~\ref{tbl:tbl1})\footnote{The asterisk
  in Tbl.~\ref{tbl:tbl1} denotes that the city is in the county of
  Genesee.}.

\renewcommand\tabcolsep{11.5pt}

\begin{table}[htbp]
\begin{tabular}{lll}
\toprule
{} &                Cities &           Counties \\
\midrule
1  &             Flint, MI* &        Genesee, MI \\
2  &           Gaylord, MI &  Dist Columbia, DC \\
3  &       Grand Blanc, MI* &         Otsego, MI \\
4  &      Mount Morris, MI* &          Wayne, MI \\
5  &  Bloomfield Hills, MI &         Ingham, MI \\
6  &           Lansing, MI &      Washtenaw, MI \\
7  &            Sedona, AZ &       Multiple, GA \\
8  &           Davison, MI* &           Kent, MI \\
9  &     Traverse City, MI &       Coconino, AZ \\
10 &         Ann Arbor, MI &           Cook, IL \\
\bottomrule
\end{tabular}
\caption{Residents interested in \#FlintWaterCrisis.}
\label{tbl:tbl1}
\end{table}

\begin{figure}[htbp]
\centering
\includegraphics[width=\columnwidth]{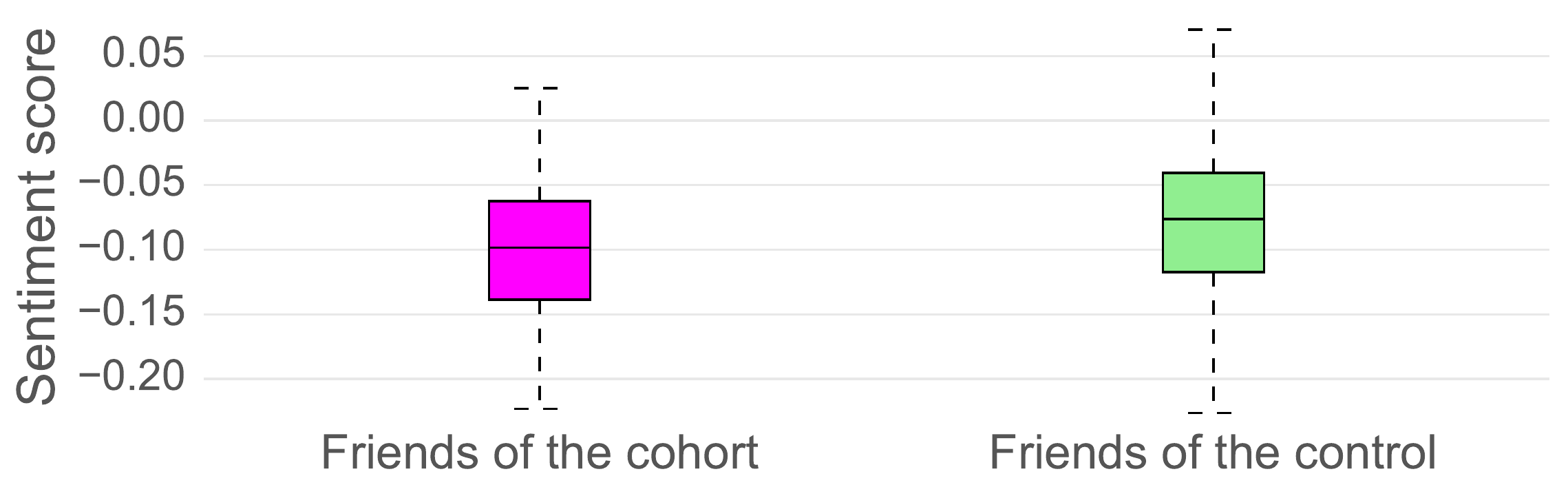}
\caption{Users and friends have similar sentiments.}\label{fig:contagion}
\end{figure}

\textbf{Contagion of Complaining.} The friends of the Flinters who expressed negative sentiments on the Flint water crisis (cohort's friends) are expected to be more negative than the friends of those Flinters who talk positively (control's friends).
Fig.~\ref{fig:contagion} illustrates that the mean sentiments of the
tweets of the cohort's friends are more negative than that of control's
friends. Furthermore, the Kolmogorov-Smirnov test statistically shows
that the distributions of the sentiment scores of tweets of the two
groups' friends are significantly different from each other with 95\%
confidence level (p-\(value\) = \(0.017\)). This discrepancy supports
our hypothesis.

\section{Related Work}\label{related-work}

Complaint is defined as ``an expression of dissatisfaction for the
purpose of drawing attention to a perceived misconduct by an
organization and for achieving personal or collective goals''
{[}\protect\hyperlink{ref-einwiller_handling_2015}{7}{]}. These goals
could be personal like ``anxiety reduction, vengeance, advice seeking,
self-enhancement'' or they could be collective such as ``helping others
and the organization''. While Einwiller and Stein
{[}\protect\hyperlink{ref-einwiller_handling_2015}{7}{]} study how large
companies (``the blamed'') handle complaints on their social media
pages, in this study we are interested in the ways citizens (``the
blamers'') raise their voice against the agents in the government as a
response to the violation of a basic human right, access to clean city
water. Rather than studying marketing or public relations aspects of the
blame process, here we are interested in the other side of the coin, how
the sociology of disasters work in this process.

By studying different kinds of crises, Oltenau et al.
{[}\protect\hyperlink{ref-olteanu_what_2015}{23}{]} categorize
information types shared on social media during these events. Following
their topology, Flint water crisis is an instantaneous human-induced
accidental hazard diffused over a county. It is a \emph{man-made
disaster} that might have started as an accident but evolved into ``a
story of government failure, intransigence, unpreparedness, delay,
inaction, and environmental injustice''
{[}\protect\hyperlink{ref-flintwater_advisory_task_force_flint_2016}{9}{]};
and it is an \emph{instantaneous crisis} because no notices were given
before it happened. Reviewing the earlier work in the literature,
{[}\protect\hyperlink{ref-olteanu_what_2015}{23}{]} identify six broad
categories for information communicated over Twitter during disasters.
These information categories are i) affected individuals, ii)
infrastructure and utilities, iii) donations and volunteers, iv) caution
and advice, v) sympathy and emotional support, and vi) other useful
information. Oltenau et al. do not consider attribution of
responsibility and blame as a distinct category; ``updates about the
investigation and suspects'' is the most related phenomenon mentioned,
which is addressed in the ``other useful information'' category
(expressed vis-à-vis shooting and bombing events).

Lin and Margolin {[}\protect\hyperlink{ref-lin_ripple_2014}{20}{]} and
Wen and Lin {[}\protect\hyperlink{ref-xu_sensing_2016}{36}{]} examine
community-level and individual-level expressions right after terrorist
attacks respectively. Both of the studies employed dictionary based
tools (SentiSense
{[}\protect\hyperlink{ref-de_albornoz_sentisense:_2012}{1}{]} and LIWC
{[}\protect\hyperlink{ref-tausczik_psychological_2010}{33}{]}) to
measure the emotions in the tweets. Lin and Margolin find that the
extent to which residents of a city visit the directly affected-city
(Boston in their case) has the most predictive power for the level of
fear, solidarity and sympathy expression in that city. Similarly,
analysis of Wen and Lin shows that ``a greater level of anxiety was
associated with locations closer to the attack site''. One of the major
differences of our research is the phase of disaster we are studying.
Instead of studying emotions right after a terrorist attack (i.e.~during
the response phase), our study focuses on political responsibility
attribution later in the recovery phase.

Regarding the methodology, Lin and Margolin
{[}\protect\hyperlink{ref-lin_ripple_2014}{20}{]} make use of two
hashtags (\#prayforboston and \#bostonstrong) as proxies for ``empathic
concern'' and ``solidarity'', to measure social support. They also adapt
keyword matching methods to detect fear and joy in the tweets, two of
the fourteen sentiments defined in the SentiSense lexicon
{[}\protect\hyperlink{ref-de_albornoz_sentisense:_2012}{1}{]}.
Similarly, Wen and Lin {[}\protect\hyperlink{ref-xu_sensing_2016}{36}{]}
measure anxiety, sadness, and anger using LIWC lexicon in French
{[}\protect\hyperlink{ref-tausczik_psychological_2010}{33}{]}. Here, we
measure sentiments of tweets using Vader sentiment analysis tool
{[}\protect\hyperlink{ref-hutto_vader:_2014}{12}{]}.

Although altruistic behaviors are common in times of disasters
{[}\protect\hyperlink{ref-glasgow_your_2016}{11}{]}, in the recovery
phase, when the community is pressed by difficult living conditions and
when there is a lack of short-term improvement, public increase their
criticism of administrators, and attribute blame to whom they perceive
as the agents of responsibility. Among other psychological and social
reasons discussed in Sec.~\ref{sec:hypotheses}, a political explanation
for this act is that in democracies the public acts as a watchdog and
actively participates in discussions to control and influence the
decision makers. To this end, use of Twitter hashtags in citizen
protests has already become a common apparatus the public leverages to
grab attention to their concerns
{[}\protect\hyperlink{ref-feenstra_democracy_2014}{8}{]}. This online
activism, sometimes called slacktivism, is defined as ``as low-risk,
low-cost activity via social media, whose purpose is to raise awareness,
produce change, or grant satisfaction to the person engaged in the
activity'' {[}\protect\hyperlink{ref-rotman_slacktivism_2011}{27}{]}. In
this study, we are not interested in how social media are used in
physical protests or how social support is expressed online, rather, we
examine theories of attribution of responsibility and blame in the
recovery phase of a crisis using observational data from social media.

\section{Conclusion}\label{conclusion}

In this study, building on the existing research in sociology of
disasters, we first form several theoretical hypotheses on attribution
of responsibility and blame. Then we operationalize these hypotheses on
unobtrusive, observational social media data via computational methods.
The findings support all of our hypotheses. However we should note that
we did not form our hypotheses upon conflicting views on the topics in
the first place, and they do not challenge the findings in the
literature. The nature of our hypotheses also does not require complex
or multivariate analysis. Yet, as acknowledged by the researchers in the
field, there is a paucity of studies on the attribution of
responsibility and blame in disaster research and a need for empirical
support. This paper adds to the few number of studies on this topic. It
contributes to the sociology of disasters research also by exploiting a
new, rarely used data source (the social web), and employing new
computational methods (e.g.~sentiment analysis and retrospective cohort
study design) on this new form of data.\footnote{Source code and data
  available at
  \href{https://github.com/oztalha/Flint}{github.com/oztalha/Flint}.} In
this regard, this work should be seen as the first step toward drawing
more challenging inferences on the sociology of disasters from social
media data.

\section{References}\label{references}

\hypertarget{refs}{}
\hypertarget{ref-de_albornoz_sentisense:_2012}{}
{[}1{]} de Albornoz, J.C., Plaza, L. and Gervás, P. 2012. SentiSense: An
easily scalable concept-based affective lexicon for sentiment analysis.

\hypertarget{ref-bartels_beyond_2002}{}
{[}2{]} Bartels, L.M. 2002. Beyond the Running Tally: Partisan Bias in
Political Perceptions. \emph{Political Behavior}. 24, 117--150.

\hypertarget{ref-bird_natural_2009}{}
{[}3{]} Bird, S., Klein, E. and Loper, E. 2009. \emph{Natural language
processing with Python}. O'Reilly Media, Inc.

\hypertarget{ref-bucher_blame_1957}{}
{[}4{]} Bucher, R. 1957. Blame and Hostility in Disaster. \emph{American
Journal of Sociology}. 62, 467--475.

\hypertarget{ref-fonger_detroit_2013}{}
{[}5{]} Detroit gives notice: It's terminating water contract covering
Flint, Genesee County in one year: 2013.
\emph{\url{http://www.mlive.com/news/flint/index.ssf/2013/04/detroit_gives_notice_its_termi.html}}.
Accessed: 2016-06-14.

\hypertarget{ref-drabek_human_1986}{}
{[}6{]} Drabek, T.E. 1986. \emph{Human System Responses to Disaster: An
Inventory of Sociological Findings}. Springer New York.

\hypertarget{ref-einwiller_handling_2015}{}
{[}7{]} Einwiller, S.A. and Steilen, S. 2015. Handling complaints on
social network sites -- An analysis of complaints and complaint
responses on Facebook and Twitter pages of large US companies.
\emph{Public Relations Review}. 41, 195--204.

\hypertarget{ref-feenstra_democracy_2014}{}
{[}8{]} Feenstra, R.A. and Casero-Ripollés, A. 2014. Democracy in the
Digital Communication Environment: A Typology Proposal of Political
Monitoring Processes. \emph{International Journal of Communication}. 8,
21.

\hypertarget{ref-flintwater_advisory_task_force_flint_2016}{}
{[}9{]} Flint Water Advisory Task Force Final Report: 2016.
\emph{\url{https://www.michigan.gov/documents/snyder/FWATF_FINAL_REPORT_21March2016_517805_7.pdf}}.
Accessed: 2016-06-16.

\hypertarget{ref-wright_genesee_2013}{}
{[}10{]} Genesee County News Release: 2013.
\emph{\url{http://media.mlive.com/newsnow_impact/other/Genesee\%20County\%20news\%20release.pdf}}.
Accessed: 2016-06-14.

\hypertarget{ref-glasgow_your_2016}{}
{[}11{]} Glasgow, K., Vitak, J., Tausczik, Y. and Fink, C. 2016. ``With
Your Help. We Begin to Heal'': Social Media Expressions of Gratitude in
the Aftermath of Disaster. \emph{Social, Cultural, and Behavioral
Modeling: 9th International Conference, SBP-BRiMS 2016, Washington, DC,
USA, June 28-July 1, 2016, Proceedings}.

\hypertarget{ref-hutto_vader:_2014}{}
{[}12{]} Hutto, C.J. and Gilbert, E. 2014. VADER: A Parsimonious
Rule-Based Model for Sentiment Analysis of Social Media Text.
\emph{Eighth International AAAI Conference on Weblogs and Social Media}.

\hypertarget{ref-iyengar_is_1994}{}
{[}13{]} Iyengar, S. 1994. \emph{Is anyone responsible?: How television
frames political issues}. University of Chicago Press.

\hypertarget{ref-kowalski_complaints_1996}{}
{[}14{]} Kowalski, R.M. and Western Carolina U 1996. Complaints and
complaining: Functions, antecedents, and consequences.
\emph{Psychological Bulletin}. 119, 179--196.

\hypertarget{ref-krugman_michigans_2016}{}
{[}15{]} Krugman, P. 2016. Michigan's Great Stink. \emph{The New York
Times}.

\hypertarget{ref-kumar_tweettracker:_2011}{}
{[}16{]} Kumar, S., Barbier, G., Abbasi, M.A. and Liu, H. 2011.
TweetTracker: An Analysis Tool for Humanitarian and Disaster Relief.
\emph{ICWSM}.

\hypertarget{ref-kwak_what_2010}{}
{[}17{]} Kwak, H., Lee, C., Park, H. and Moon, S. 2010. What is Twitter,
a Social Network or a News Media? \emph{Proceedings of the 19th
International Conference on World Wide Web}, New York, NY, USA.

\hypertarget{ref-landis_measurement_1977}{}
{[}18{]} Landis, J.R. and Koch, G.G. 1977. The Measurement of Observer
Agreement for Categorical Data. \emph{Biometrics}. 33, 159--174.

\hypertarget{ref-lau_two_1985}{}
{[}19{]} Lau, R.R. 1985. Two Explanations for Negativity Effects in
Political Behavior. \emph{American Journal of Political Science}. 29,
119--138.

\hypertarget{ref-lin_ripple_2014}{}
{[}20{]} Lin, Y.-R. and Margolin, D. 2014. The ripple of fear, sympathy
and solidarity during the Boston bombings. \emph{EPJ Data Science}. 3.

\hypertarget{ref-mcpherson_birds_2001}{}
{[}21{]} McPherson, M., Smith-Lovin, L. and Cook, J.M. 2001. Birds of a
Feather: Homophily in Social Networks. \emph{Annual Review of
Sociology}. 27, 415--444.

\hypertarget{ref-neal_blame_1984}{}
{[}22{]} Neal, D.M. 1984. Blame Assignment in a Diffuse Disaster
Situation: A Case Example of the Role of an Emergent Citizen Group.
\emph{International journal of mass emergencies and disasters}. 2,
251--266.

\hypertarget{ref-olteanu_what_2015}{}
{[}23{]} Olteanu, A., Vieweg, S. and Castillo, C. 2015. What to Expect
When the Unexpected Happens: Social Media Communications Across Crises.
\emph{Proceedings of the 18th ACM Conference on Computer Supported
Cooperative Work \& Social Computing}, New York, NY, USA.

\hypertarget{ref-palen_crisis_2007}{}
{[}24{]} Palen, L., Vieweg, S., Sutton, J., Liu, S.B. and Hughes, A.L.
2007. Crisis informatics: Studying crisis in a networked world.
\emph{Proceedings of the Third International Conference on E-Social
Science}.

\hypertarget{ref-the_white_house_president_2016}{}
{[}25{]} President Obama Signs Michigan Emergency Declaration: 2016.
\emph{\url{https://www.whitehouse.gov/the-press-office/2016/01/16/president-obama-signs-michigan-emergency-declaration}}.
Accessed: 2016-06-02.

\hypertarget{ref-longley_report:_2011}{}
{[}26{]} Report: Buying in to new water pipeline from Lake Huron cheaper
for Flint drinking water than treating river water: 2011.
\emph{\url{http://www.mlive.com/news/flint/index.ssf/2011/09/water_treatment.html}}.
Accessed: 2016-06-14.

\hypertarget{ref-rotman_slacktivism_2011}{}
{[}27{]} Rotman, D., Vieweg, S., Yardi, S., Chi, E., Preece, J.,
Shneiderman, B., Pirolli, P. and Glaisyer, T. 2011. From Slacktivism to
Activism: Participatory Culture in the Age of Social Media. \emph{CHI
'11 Extended Abstracts on Human Factors in Computing Systems}, New York,
NY, USA.

\hypertarget{ref-sears_selective_1967}{}
{[}28{]} Sears, D.O. and Freedman, J.L. 1967. Selective Exposure to
Information: A Critical Review. \emph{The Public Opinion Quarterly}. 31,
194--213.

\hypertarget{ref-shear_ive_2016}{}
{[}29{]} Shear, M.D. and Bosman, J. 2016. ``I've Got Your Back,'' Obama
Tells Flint Residents. \emph{The New York Times}.

\hypertarget{ref-simon_politicized_2001}{}
{[}30{]} Simon, B. and Klandermans, B. 2001. Politicized collective
identity: A social psychological analysis. \emph{American Psychologist}.
56, 319--331.

\hypertarget{ref-singer_introduction_1982}{}
{[}31{]} Singer, T.J. 1982. An introduction to disaster: Some
considerations of a psychological nature. \emph{Aviation, space, and
environmental medicine}.

\hypertarget{ref-spence_social_2016}{}
{[}32{]} Spence, P.R., Lachlan, K.A. and Rainear, A.M. 2016. Social
media and crisis research: Data collection and directions.
\emph{Computers in Human Behavior}. 54, 667--672.

\hypertarget{ref-tausczik_psychological_2010}{}
{[}33{]} Tausczik, Y.R. and Pennebaker, J.W. 2010. The Psychological
Meaning of Words: LIWC and Computerized Text Analysis Methods.
\emph{Journal of Language and Social Psychology}. 29, 24--54.

\hypertarget{ref-tobler_computer_1970}{}
{[}34{]} Tobler, W.R. 1970. A Computer Movie Simulating Urban Growth in
the Detroit Region. \emph{Economic Geography}. 46, 234--240.

\hypertarget{ref-wallace_human_1956}{}
{[}35{]} Wallace, A.F.C. 1956. \emph{Human behavior in extreme
situations; a study of the literature and suggestions for further
research}. Washington, National Academy of Sciences, National Research
Council.

\hypertarget{ref-xu_sensing_2016}{}
{[}36{]} Wen, X. and Lin, Y.-R. 2016. Sensing Distress Following a
Terrorist Event. \emph{Social, Cultural, and Behavioral Modeling}. K.S.
Xu, D. Reitter, D. Lee, and N. Osgood, eds. Springer International
Publishing. 377--388.

\end{document}